\documentstyle[12pt]{article}
\def\lromn#1{\uppercase\expandafter{\romannumeral#1}}

\begin{document}

\begin{flushright}
TU/96/495
\end{flushright}

\vspace{12pt}

\begin{center}
\begin{Large}

\renewcommand{\thefootnote}{\fnsymbol{footnote}}
\bf{
Particle Production and Inflaton Decay
}
\footnote[1]
{To appear in the Proceedings of the Symposium on 
"The Cosmological Constant and the Evolution of the Universe",
Universal Academic Press (Tokyo),
held November 7-10, 1995.
}

\end{Large}

\vspace{36pt}

\begin{large}
M. Yoshimura \\

Department of Physics, Tohoku University\\
Sendai 980-77 Japan\\
\end{large}

\vspace{54pt}

{\bf ABSTRACT}
\end{center}

\vspace{0.5cm} 
A new theory of thermal history after inflation is outlined
by taking into account the effect of parametric resonance caused by
inflaton oscillation. A typical outcome is existence of
a high temperature phase, that makes possible GUT baryogenesis
consistently with low enough gravitino abundance.

%%%%%%%%%%%%%%%%%%%%%%%%%%%%%%%%%%%%%%%%%%%%%%%%%%%%%
\newpage
It is natural, and even inevitable as the fundamental microscopic theory
advances, to ask how things in the universe are as they are.
The classic example in cosmology is how the element abundance is as it is;
the problem of nucleosynthesis.
In recent times we went quite further along these lines:
we now even talk about origin of matter-antimatter asymmetry, creation
of the spacetime itself, and so on.
What I would like to discuss in this talk is how the bulk of entropy
in our universe, the universe at least in the observable portion to us,
has been created. 
The problem might sound purely accademic, but actually 
has very practical aspects with regard to
the problem of baryogenesis at the Planck or the GUT epoch, as explained
later.

We shall discuss the entropy generation in the paradigm of inflationary
universe models. This puts our task to the logical extreme, since
at the end of the exponential expansion of inflationary stage
our portion of universe becomes radiation free:
what is left from inflation is simply oscillation of the inflaton field 
(coherent and homogeneous field that causes inflation)
around the minimum of inflaton potential.
Thus from the entropy point there is essentially nothing at the start.
We must understand how the entire entropy of our universe is generated
originating from inflaton oscillation.
This problem has been often called reheating problem, although it is 
somewhat misleading to put the word "re"
because it does not matter whether there is an epoch
of thermal equilibrium prior to inflation.
The elementary picture of reheating until the middle of 1994 was as follows 
\cite{reheat-original}:
the coherent and homogeneous inflaton oscillation can be regarded as
an aggregate of particles at rest with a precisely tuned phase relation
and these nonrelativistic particles independently
decay according to the probability law given by the simple Born decay formula.
Assuming these produced particles immediately
interact strongly, one may hope to 
achieve rapid thermalization among created particles.

Recent works on this subject made it clear 
that this elementary picture is grossly wrong if the
initial oscillation amplitude of the inflaton field is large
\cite{reheating parametric}, \cite{holman 95}, 
\cite{linde et al 94}, \cite{brandenberger et al}, \cite{mine95-1},
\cite{fkyy95-1}, \cite{fkyy95-2}.
Particle production under a large amplitude and almost periodic oscillation
is dictated by many instability bands only one of which gives the
Born decay rate, a valid approximation under the small, but not under
the large amplitude limit.
Existence of infinitely many bands under a periodic coefficient function
is known as the phenomenon of parametric resonance \cite{landau-lifschitz m},
\cite{coddington}.
Incorporation of this new source of many instability bands gives rise to much
faster inflaton decay and much more copious particle production.
Thus the thermal history right after inflation is quite different from
what was anticipated previously: there exists a prolonged epoch of
much higher temperatures than expected from the Born formula
\cite{fkyy95-2}.
It opens up a new possibility of baryogenesis consistently with
the gravitino problem, as discussed later.
A different picture of the reheating has been advocated by Linde
in the preceeding talk \cite{linde et al 95}.

As a model of inflation we take the simplest; chaotic inflation with
a mass term potential 
\( \:
\frac{1}{2}\, m_{\xi }^{2}\,\xi ^{2}
\: \). At the Hubble rate equal to the inflaton mass,
damped oscillation starts. To be consistent with COBE anisotropy
of the microwave fluctuation we may take the inflaton mass in the range of
\( \:
m_{\xi } = 10^{13} \,{\rm GeV}.
\: \)
The inflaton field has coupling to ordinary matter and gauge fields, which
we generically denote by $\varphi $. Only bose fields are relevant to
the following discussion.
These couplings may be either of the quartic $\xi ^{2}\varphi ^{2}$ or of
the Yukawa $\xi \varphi ^{2}$ type. Ratio of these two terms is an important
quantity to the rest of discussion, 
but without much guidance from the fundamental theory
we take a simple supersymmetric model with a superpotential of the form,
\( \:
{\cal W} = \frac{1}{2}\, m_{\xi }\xi ^{2} + \frac{1}{\sqrt{2}}\,
g\xi \varphi ^{2},
\: \)
which gives an ordinary potential for the $\xi $ coupling,
\( \:
V_{\xi } = \frac{1}{2}\, g^{2}\xi ^{2}\varphi ^{2} + \frac{1}{2}\, 
gm_{\xi }\xi \varphi ^{2} \,.
\: \)
Fixing the ratio of the quartic to the Yukawa coupling terms
to roughly of order $g\xi /m_{\xi }$ is important, but
a precise ratio number given here does not matter to the rest of conclusions.

For the time being, let us ignore time variation of the inflaton
amplitude due to the cosmological expansion. 
We find it best to exploit the Schr$\stackrel{..}{{\rm o}}$dinger picture
\cite{mine95-1}
in order to clarify the nature of quantum state under the periodic
perturbation of the inflaton field. In this formalism any quantum state
can be expressed in terms of a direct product of the state vectors
$|\,\psi (t)\,\rangle_{k} $ which obeys the Schr$\stackrel{..}{{\rm o}}$dinger
equation of a periodic frequency $\omega^{2}_{k} (t)$.
It turns out \cite{mine95-1} 
that the quantum wave function is solved with a Gaussian ansatz;
\begin{eqnarray*}
\langle q_{k}|\,\psi (t)\,\rangle_{k} =
\frac{1}{\sqrt{|u_{k}(t)|}}\,\exp [\,-\,\frac{\pi }{2|u_{k}(t)|^{2}}\,q_{k}^{2}
+ \frac{i}{4}\frac{d}{dt}\,\ln |u_{k}(t)|^{2}\cdot q_{k}^{2}\,] \,,
\end{eqnarray*}
where the time dependent function $u_{k}(t)$ obeys 
the same classical oscillator equation as the mode equation
corresponding to the wave vector $\vec{k}$;
\begin{eqnarray*}
\frac{d^{2}u _{k}}{dt^{2}} + \omega_{k} ^{2}(t)\,u _{k} = 0 \,, 
\hspace{0.5cm} 
\omega_{k} ^{2}(t) = \vec{k}^{2} + g^{2}\xi ^{2}+ gm_{\xi }\xi \,, 
\end{eqnarray*}
with
\( \:
\xi = \xi _{0}\,\cos [\,m_{\xi }(t - t_{0})\,] \,.
\: \)
Moreover, an initial choice of the quantum state can be translated to
a choice of the boundary condition to this classical equation.
We took the ground state since inflation pushes out
essentially all matter out of horizon. 
The initial vacuum at $t = t_{0}$ then corresponds to
\( \:
u(t) \sim (\omega (t_{0})/\pi )^{-1/2}\,e^{i\omega (t_{0})\,(t-t_{0})} \,.
\: \)
An important point here is that there is no freedom to impose the
boundedness at $t = \infty $, quite unlike the Bloch wave solution
in condensed matter physics (time is replaced by coordinate there).

A mathematical theorem \cite{coddington} 
tells that with periodicity of $\omega ^{2}(t)$ alone,
and irrespective of any details of this function, there are infinitely
many bands of instability separated by stability bands
in the two parameter space that characterizes $\omega ^{2}(t)$,
$k$ and the oscillation amplitude $\xi _{0}$.
Instability grows exponentially with time: the asymptotic form of solution
is 
\( \:
u(t) \:\rightarrow  \: e^{\mu t}\times 
\: \)
(periodic function) with $\mu > 0$.
Furthermore, within each instability band denoted by $n$
there is an eigenmode along a curve in the parameter space,
\( \:
k = k_{n}(\xi_{0} ) \,, \; (n = 1 \,, 2\,, 3\,,\cdots ),
\: \)
that
has periodic zeros at $t_{i} \: (i = 1 \,, 2 \,, 3 \,, \cdots )$, 
$u_{n}(t_{i}) = 0$. 
This eigenmode approaches 
$k \:\rightarrow  \:\frac{n}{2}\,m_{\xi }$ as $\xi_{0} \:\rightarrow  \: 0$.
This remarkable fact has a profound implication on the behavior of 
the quantum state: the phase factor 
\( \:
\frac{d}{dt}\,\ln |u(t)|^{2}
\: \)
in the wave function $\langle q|\,\psi (t)\,\rangle $ diverges at these 
$t_{i}$.
All off-diagonal density matrix elements
\( \:
\rho (q \,, q' \,, t) = \langle q|\,\psi (t)\,\rangle \,
\langle q'|\,\psi (t)\,\rangle^{*} \,, \;(q \neq q') \,, 
\: \)
vanish, when time averaged over short time scales of order $1/m_{\xi }$.
We take this as a signal to indicate an almost classical behavior in the
quantum system under the periodic perturbation.
The coarse grained density matrix thus defined is diagonal and positive
definite: in the harmonic oscillator Fock base of
\( \:
H_{\omega }\,|n\rangle = \omega \,(n+\frac{1}{2})|n \rangle  \,, \;
(n = 0 \,, 1\,, 2 \,, \cdots ) \,, \;
\rho _{n\,, m} = \rho ^{D}_{n}\,\delta _{nm} \,,
\: \)
with
\begin{eqnarray*}
\sum_{n = 0}^{\infty }\,\rho ^{D}_{n} = 1 \,, \hspace{0.5cm} 
\sum_{n = 0}^{\infty }\,(\rho ^{D}_{n})^{2} < 1
\,.
\end{eqnarray*}
Thus the system is effectively described by a classical probability
distribution function $\rho ^{D}$.

This classical instability implies that the wave function of 
the quantum system has an exponentially growing Gaussian width
\( \:
\langle q^{2} \rangle \propto |u(t)|^{2}
\: \), which
means that higher energy levels are excited with time, 
to be interpreted as particle production in the particle language. 
In this way one can precisely compute
produced number of particles in terms of the density matrix element,
resulting in
\( \:
\langle N \rangle = \rho _{00}^{-2} - 1
\: \)
for each mode. The formalism manifestly respects unitarity, and
one may devise useful approximation for many quantities.
For instance, the particle number distribution $\rho _{2n\,, 2n}$ is
expressed in terms of the average number of particles $\langle N \rangle$
alone, which as $n \:\rightarrow  \:\infty $, has an asymptotic form of
\begin{eqnarray*}
\rho _{2n\,, 2n} = \rho ^{D}_{2n\,, 2n} \:\rightarrow  \:
\frac{1}{\sqrt{\pi \,(n+1)}}\,\frac{e^{-n/\langle N \rangle}}
{\sqrt{\langle N \rangle}} \,.
\end{eqnarray*}
I would like to stress that this distribution is rather broad, as evidenced
for instance by the calculated dispersion 
\( \:
\sqrt{\langle N^{2}\rangle - \langle N \rangle^{2}}
/\langle N \rangle \sim 1 \,.
\: \)
The pseudo-entropy defined by
\( \:
-\,{\rm tr}\,\rho ^{D}\ln \rho ^{D}
\: \)
is calculated as
\( \:
\sim \ln \langle N \rangle + 0.09 
\: \)
using the asymptotic formula,
which grows logarithmically with the degrees of freedom 
$\langle N \rangle$ as usual.

With cosmological expansion one has to consider damping of the inflaton
amplitude $\xi (t)$. 
For the Hubble time $1/H$
much larger than the oscillation time scale $1/m_{\xi }$ one may
deal with averaged quantities over one oscillatin time.
In the rest of discussion we take $\xi $ to mean the time-averaged amplitude.
In the model of our concern we may separate the initial and the final phase
of inflaton oscillation, depending on the value of
\( \:
\eta \equiv g\xi /m_{\xi } \,.
\: \)
At the very start of the $\xi $ oscillation this value is of order
\( \:
\eta_{0} = g\xi _{0}/m_{\xi } \approx 10^{6}\,g \,, 
\: \)
which is large unless $g$ is extremely small.
Both the large $\eta $ and the small $\eta $ regions are approximately
described by the standard Mathieu equation;
\begin{eqnarray*}
\frac{d^{2}u}{d\tau ^{2}} + (\,h - 2\theta \cos (2\tau )\,)u = 0 \,.
\end{eqnarray*}
For instance, in the large $\eta $ limit initially relevant, the quartic
inflaton coupling dominates and
\( \:
h = (\vec{k}^{2} + \frac{1}{2}\, g^{2}\xi ^{2})/m_{\xi }^{2} \,, \;
\theta = g^{2}\xi ^{2}/(4m_{\xi }^{2}) = \eta ^{2}/4
\,, \; \tau = m_{\xi }t \,.
\: \)
Thus the relevant instability band is limited to
\( \:
h - 2\theta \,(= \vec{k}^{2}/m_{\xi }^{2}) > 0 \,.
\: \)
Although much is known on instability of the Mathieu equation
\cite{mathieu eq}, 
we had to improve much more in the large and the small $\theta $ region,
especially on analytic form of the growth rate $\lambda (h \,, \theta )$
in the late time behavior of $e^{\lambda m_{\xi }t}$.
Detailed results of this investigation are left to Ref 
\cite{fkyy95-1}, \cite{fkyy95-2}, but let me mention a few key points.

First, in the large amplitude limit one may use analogy to scattering
problem in quantum mechanics, to analyze the growth rate in instability
bands. Starting at the initial "position", $t_{0}$, one considers
infinitely many barriers of the same height, each of which can be replaced by
an inverted harmonic oscillator in the large $\theta $ limit.
Only the growing component is amplified each time of the barrier crossing,
and one may obtain an analytic formula for the growth rate
by diagonalizing the transfer matrix over one barrier crossing. 
The result is summarized \cite{fkyy95-1} as follows: with
\( \:
\epsilon \equiv \frac{h}{2\theta } - 1 \,, 
\: \)
\begin{eqnarray}
&&
\lambda = \frac{1}{\pi }\,\ln (\,\sqrt{x} + \sqrt{x-1}\,) \,, \hspace{0.5cm} 
x = (1 + e^{-\,\pi \sqrt{\theta }\,\epsilon })\,\cos ^{2}\psi \,, 
\nonumber 
\\
&&
\psi = \frac{\pi ^{2}}{2}\,\sqrt{\theta } + \sqrt{\theta }\,\epsilon 
\ln (\pi\,\theta ^{1/4}) +
\frac{1}{2}\, \Im \,\ln [\,
\frac{\Gamma \left( \,(1 - i\sqrt{\theta }\epsilon )/2\,\right)}
{\Gamma \left( \,(1 + i\sqrt{\theta }\epsilon )/2\,\right)}\,] \,.
\nonumber 
\end{eqnarray}
The instability island that appears as $\epsilon $ varies with a fixed
$\theta $ is characterized by $x>1$, separated by stability
regions of $x<1$.
The maximal rate $\lambda $ is $\sim 0.28$ at $\epsilon = 0$, 
and as $\epsilon $ increases
from 0, the rate decreases. The averaged value taken over a large region
of $\theta $ is $\approx 0.15$ for the first island of smallest
$\epsilon $. Since it is difficult to deal with each instability
band separately and then add them together, we use this average rate
and multiply the phase space volume of
\begin{eqnarray*}
O[\,\frac{(gm_{\xi }\xi )^{3/2}}{2\pi ^{2}}\,]
\end{eqnarray*}
contributing to this rate.
The average energy of created $\varphi $ particle is of order
\( \:
\sqrt{gm_{\xi }\xi } \,,
\: \)
which is initially very large
\( \:
\approx \sqrt{g}\cdot 10^{16}\,{\rm GeV} \,.
\: \)

On the other hand, in the the small amplitude limit the Yukawa coupling term
dominates and the Mathieu parameter becomes
\( \:
h = 4\vec{k}^{2}/m_{\xi }^{2} \,, \; \theta = 2g\xi /m_{\xi } = 2\eta \,.
\: \)
The decay of the quantum ground state is described by the exponential
law, $\rho _{00} \approx e^{-\Gamma Vt}$ with $\Gamma $ 
the decay rate per unit volume and
per unit time. It consists of contribution from each band $\Gamma _{n}$;
$\Gamma  = \sum_{n = 1}^{\infty }\,\Gamma _{n}$.
Precise formula from each band is given in Ref \cite{mine95-1}, and goes like
\begin{eqnarray*}
\Gamma _{n} \,\propto\, 
m_{\xi }^{4}\sqrt{1 - \frac{4m^{2}}{n^{2}m_{\xi }^{2}}}
\,(\frac{g\xi }{m_{\xi }})^{2n} \,.
\end{eqnarray*}
For clarity we included the mass $m$ dependence of $\varphi $ field.
This formula implies that the process from the $n-$th band is a multi-particle
reaction of $n\,\xi \:\rightarrow  \: 2\,\varphi $ with $\varphi $ particle
energy of $nm_{\xi }/2$.
The most obvious contribution is from the first band, for which
\( \:
\Gamma _{1} = g^{2}m_{\xi }^{2}\,\xi ^{2}/(64\pi ) = n_{\xi }\,\gamma _{\xi }
\: \)
with $\gamma _{\xi }$ one particle decay rate of the inflaton 
$(\xi \,\rightarrow  \, \varphi \,\varphi) $ and
$n_{\xi } = m_{\xi }\,\xi ^{2}/2$ the inflaton number density.
This relation confirms the expected result that the leading effect in
the small $\xi $ amplitude limit is the stochastic 2-body decay.

Particle production accompanies back reaction against the inflaton and
ultimately the inflaton oscillation ends by damping due to particle 
production. For a global consideration
this back reaction can be treated \cite{fkyy95-2} 
via energy balance between the inflaton
and the created particle energy. With cosmological expansion included,
time evolution equation is given by 
\begin{eqnarray}
&&
\frac{d\rho _{\xi }}{dt} + 3H\rho _{\xi } = - 
\,\frac{d}{dt}\,N\,\langle \rho _{\varphi } \rangle - 
\gamma _{\xi }\,\rho _{\xi } \,, 
\nonumber \\
&&
\frac{d\rho _{r}}{dt} + 4H\rho _{r} = 
\,\frac{d}{dt}\,N\,\langle \rho _{\varphi } \rangle +
\gamma _{\xi }\,\rho _{\xi } \,, 
\nonumber \\
&&
H^{2} = \frac{8\pi }{3}G\,(\rho _{\xi }+ \rho _{r}) \,, 
\nonumber  \\
&&
\frac{dm_{\xi }^{2}}{dt} = g^{2}\,\frac{d}{dt}\,\langle \varphi ^{2} \rangle
\,. \nonumber 
\end{eqnarray}
Here $N$ is a number of bose fields contributing to the parametric
resonance, $8$ in the minimal supersymmetric model in which 2 Higgs
doublets couple to the inflaton field.
We also considered fluctuation 
\( \:
\langle \varphi ^{2} \rangle
\: \)
associated with particle production (fluctuation-dissipation theorem), 
causing the $\xi $ mass variation.
It turns out \cite{fkyy95-2} that the source term on the right hand side
\( \:
\frac{d}{dt}\,\langle \rho _{\varphi } \rangle
\: \)
can be handled without invoking integro-differential equations both in
the large and the small amplitude limit. The remaining complexity is how to
match these two regions. We used two types of matching, one crude and
the other somewhat elaborate. Result is not very sensitive to these
details of the matching. 
To understand essential features of our result,
it is sufficient to consider the simpler matching, based on
the quartic inflaton dominance until 
\( \:
\theta _{Y} \equiv 2g\xi /m_{\xi } = \sqrt{152} -12 \sim 0.33
\: \)
and the subsequent Yukawa dominance.

It is now appropriate to discuss consequences \cite{fkyy95-2} 
that come out of numerical
integration of evolution equation. We mainly studied this under the initial
 $\xi $ dominance;
\( \:
\rho _{\xi }(t_{0}) = \frac{1}{2}\, m_{\xi }^{2}\,\xi _{0}^{2} \,, \;
\rho _{r} (t_{0}) = 0 \,.
\: \)
Evolution equation was integrated until almost complete Born decay, 
conveniently taken at
\( \:
t = t_{e} = 10\,\gamma _{\xi }^{-1} \,.
\: \)
What typically occurs is that after $O[10 - 100]\times m_{\xi }^{-1}$ 
oscillations catastrophic particle production terminates the initial phase
of dissipative $\xi $ evolution
and subsequently enters into a much slower phase under the small 
amplitude oscillation, ultimately ending with the Born decay.
It can be checked \cite{fkyy95-2} 
that strong interaction leading to thermalization
occurs immediately after the catastrophic particle production.
The effect of parametric resonance in the large amplitude regime
is found most significant for a range of couplings,
\( \:
g = 10^{-4} - 0.1 \,.
\: \)
In this range of coupling the initial thermal temperature denoted by
$T_{i}$ is much higher than what the Born rate would suggest;
\( \:
T_{i} = O[10^{14} - 10^{15}]
\: \)
GeV. This high temperature phase is followed by the much slower phase of
the Born decay, until the final temperature becomes of order
\( \:
10^{14}\,g\,{\rm GeV} \,,
\: \)
at the time $t_{e}$.
This agrees, remarkably well, with the naive estimate assuming 
instantaneous decay and instantaneous thermalization due to the Born decay,
\( \:
T_{B} \sim  0.1\,\sqrt{\gamma _{\xi }m_{{\rm pl}}} \approx  10^{-2}\,g
\sqrt{m_{\xi }m_{{\rm pl}}} \,.
\: \)
But it should be noted that this simple estimate completely ignores
the parametric resonance effect, which predicts an epoch of much higher
temperatures: the thermal history after inflation cannot be described
by a single reheat temperature such as $T_{B}$.

Existence of the high temperature phase necessitates to reconsider
high energy processes such as baryogenesis and gravitino production.
Indeed, what we found in \cite{fkyy95-2} 
is that high enough initial temperature favored by GUT baryogenesis is realized
even with low enough gravitino abundance consistent with nucleosynthesis.
The old formula of the gravitino abundance
\( \:
n_{3/2}/s \sim 10^{-2}\,\frac{T_{B}}{m_{{\rm pl}}}
\: \)
with $s$ the entropy density is not valid since the thermal history
right after the catastrophic particle production is complicated in
some region of the coupling $g$ and $T_{B}$ is not a good measure to
characterize the thermal history.
Let us then consider gravitino production after inflation, following 
evolution equation for the gravitino number density $n_{3/2}$,
\begin{eqnarray*}
\frac{dn_{3/2}}{dt} + 3H\,n_{3/2} = \langle \Sigma v \rangle\,n_{\varphi }
^{2} \,,
\end{eqnarray*}
where $n_{\varphi }$ is the thermal number density of one species of
created particles.
The cross section $\langle \Sigma v \rangle$
of gravitino production $ \varphi \,\varphi 
\:\rightarrow  \:g_{3/2}\:g_{3/2}$ is different
\cite{gravitino recent}, whether or not the gravitino
is lightest supersymmetric particle (LSP). 
For the heavy gravitino $\langle \Sigma v \rangle \sim 250/m_{{\rm pl}}^{2}$.
Both possibilities of stable and unstable gravitino remain viable. 
Let us only mention a possibility
of the gravitino dominated universe at the present epoch.
With the initial temperature $T_{i} > 2\times 10^{14}\,{\rm GeV}$ imposed
to give a favorable situation for GUT baryogenesis,
there is a region of parameters for the closure density of gravitino
dominated universe if $m_{3/2} = 0.1 - 10\, {\rm GeV}$.
The basic reason this becomes possible is that the initial large gravitino
yield created right after the catastrophic particle production
is much diluted via the late phase of Born decay.
This is again a reflection of the large disparity of the two temperatures;
\( \:
T_{i} \gg T_{B} \,.
\: \)
Of course, it remains to demonstrate a sizable baryon to photon ratio.
But things are not bad: there is an epoch immediately before the catastrophic
particle production in which non-equilibrium environment necessary for
baryon generation exists, and moreover the baryon to photon ratio
is of order $10^{-10}$ allowing some amount of dilution in later epochs.

%\vspace{0.5cm} 

\end{document}